\documentclass[12pt]{article}
\usepackage{graphicx}
\usepackage{amsmath}
\usepackage{bm}
\def\Vec#1{\mbox{\boldmath $#1$}}
\def\mC{\mathcal C}

\def\itmb{\begin{itemize}}
\def\itme{\end{itemize}}
\def\enmb{\begin{enumerate}}
\def\enme{\end{enumerate}}
\def\eqnb{\begin{equation}}
\def\eqne{\end{equation}}

\title{ON THE AMPLITUDE OF EXTERNAL PERTURBATION AND\\
THE CHAOS VIA DEVIL'S STAIRCASE\\
- STABILITY OF ATTRACTORS -
}

\author{ Sadataka FURUI\\
 Graduate School of Science and Engineering, Teikyo University, \\
2-17-12 Toyosatodai, Utsunomiya, 320-0003, Japan{\thanks
{\textit{E-mail address:} furui@umb.teikyo-u.ac.jp}}\\
Tomoyuki TAKANO\\
IIX inc, Location 7F Daiichi Life Ins. Bld. Annex, \\
2-7-12 Nishi-Gotanda, Shinagawa-ku, Tokyo 141-0031, Japan{\thanks
{\textit{E-mail address:} takano\_t@iix.co.jp}}
}
\begin{document}
\maketitle

\begin{abstract}
We made a chaotic memristive circuit proposed by Chua and Muthuswamy-Chua, and analyzed the behavior of the voltage of the capacitor, electric current in the inductor and the voltage of the memristor by adding an external sinusoidal oscillation  of a type $\gamma\omega \cos\omega t$ to ${\dot y}(t)\simeq {\dot i_L}(t)$, when the ${\dot x}(t)\simeq {\dot v_C}(t)$ is expressed by $y(t)/C$, and studied Devil's staircase route to chaos. 
 
We compared the frequency of the driving oscillation $f_s$ and the frequency of the response $f_d$ in the window and assigned $W=f_s/f_d$ to each window. 
When capacitor $C=1.0$, we observe stable attractors of  Farey sequences $\displaystyle W=\{\frac{1}{2}, \frac{2}{3},\frac{3}{4},\frac{4}{5}, \cdots ,\frac{14}{15},\frac{1}{1}\}$, which can be interpreted as hidden attractors, while when $C=1.2$, we observe disturbed attractors.

Frequency dependent stability of chaotic circuits due to octonions is discussed. 
\end{abstract}

\begin{center}{ Keywords : Devil's staircase, Chaos, Octonions, Chua's circuit, Memristor}
\end{center}

\section{Introduction}
\noindent In 1993, Hayes, Grebogi and Ott\cite{HGO93} discussed an electronic circuit in a chaotic regime, 
using the characterization of chaos as deterministic noise\cite{EckRu85}.
Typical evolution equation of the physical system is characterized by
\begin{equation}
\dot {\Vec x}(t)={\Vec F}_\mu({\Vec x}(t))
\end{equation}
where $\Vec x$ is a set of coordinates in $R^m$ and $F_\mu({\Vec x})$ determines
the nonlinear time evolution. 
 The nature of asymptotic motion depends upon the parameter $\mu$ and values at which the change of asymptotic regime happens are called bifurcation points. The asymptotic motion is settled down onto an attractor in phase space, but its structure is not well understood, and  unexpected instability of attractors are observed.
Chaos is characterized by sensitivity to the initial condition, or the dependence on the difference of orbits 
\begin{equation}
d(t)=|{\Vec s}'(t)-{\Vec s}(t)|\sim e^{\lambda_1 t}.
\end{equation}
The parameter $\lambda_1$ is called the largest Lyapunov coefficient.

How to measure the strangeness and the dimension of attractors was discussed by Grassberger and Procaccia\cite{GrPr82}. 
In 1984, Grebogi et al.\cite{GOPY84} showed that there are strange attractors whose Lyapunov exponents are negative. They showed that some attractors become nonstrange by certain weak perturbations, and the structure of Strange Nonchaotic Attractor (SNA) is rather complicated. 
 In 1971,  Chua proposed that in addition to resistor, inductor and
conductor, there is an interesting non-linear circuit element which is called
memristor, a combination of memory and resistor \cite{Chua71}.
 Chua and  Kang represented the dynamical system, which is called memristive system, by
\begin{eqnarray}
\dot x&=&f(x,u,t)\nonumber\\ 
y&=&g(x,u,t)u   
\end{eqnarray}
where $u$ and $y$ denote the input and the output of the system, respectively, and $x$ denotes the state of the system\cite{ChKa76}.
Muthuswamy and Chua\cite{MuCh10} showed that one can produce a non linear memristor, and combined with  inductor, capacitor and opeamp, presented chaotic behaviors. The chaotic behavior of memristic system was analyzed in \cite{GiLe09, GGJLF10,GLC10} and \cite{Chua11}. 

Chaos in electronic circuits that consists of capacitors, inductors, resistors and opeamp was studied in 1986 in \cite{CKM86}.  When an external current source is added to the Chua's circuit, a Farey sequence period adding law was observed in \cite{PZC94}. 
Analyses of driven Chua's circuit were done in \cite{ZhLi97, BaCa98} and in \cite{ThLa02}. 
We analyzed the chaotic behaviors of Chua's circuit by adding an external current in the system\cite{FuNi07}. 
In this paper, we add an external oscillation of current to the memristic system, which we call memristive circuit, and study changes in the chaotic behaviors due to an addition of driving currents. 

Electrons are described by spinors and in the spinor theory of Cartan\cite{Cartan66}, self-dual vector field $X_i  (i=1,2,3,4)$ is produced from spinors as a Pl\"uckner coordinate. Spinors are described by quaternions $\mathcal H$ and bispinors that consist of two quaternions with a new imaginary unit $l$ composes an octonion,
\[
{\mathcal O}={\mathcal H}+ l {\mathcal H}
\]
which has the automorphism: $G_2$ group.
There are 24 dimensional bases in the $G_2$ group, and it has the triality symmetry. We study the electro magnetic field of the memristor using the octonion bases\cite{SF12,SF13a,SF13b}. 

The paper is organized as follows. In Sect.2, we present chaos in Chua's circuit, and in Sect.3 we present Chaos in Memristic circuit. In Sect.4, Strange nonchaotic attractor in driven memristor is studied, and in Sect.5,
the resonance frequency of the driven memristive circuit is analyzed by using octonion basis.
Difference of strange chaotic attractors and strange nonchaotic attractor (SNA) are discussed in Sect.6.

\section{Chaos in Chua's circuit}
Chua's circuit consists of an autonomous circuit which contains a three-segments piecewise-linear resistor, two capacitors, one inductor and a variable resistor \cite{CKM86}.
The equation of the circuit is described by the coupled differential equation
\begin{equation}\label{chua_bare}
\left\{\begin{array}{l}
C_1\frac{d}{dt}v_{C1}=G(v_{C2}-v_{C1})-\tilde g(v_{C1},m_0,m_1)\\
C_2\frac{d}{dt}v_{C2}=G(v_{C1}-v_{C2})+i_L\\
L\frac{d}{dt}i_L=-v_{C2}
\end{array}\right.
\end{equation}

 The three-segment piecewise-linear resistor which constitutes the non-linear element is characterized by
\[
\tilde g(v_{C1},m_0,m_1)=(m_1-m_0)(|v_{C1}+B_p|-|v_{C1}-B_p|)/2+m_0 v_{C1}, 
\]
where $B_p$ is chosen to be 1V, $m_0$ is the slope (mA/V) outside $|v_{C1}|>B_p$  and $m_1$ is the slope inside $-B_p<v_{C1}<B_p$.  

Here $v_{C1}$ and $v_{C2}$ are the voltages of the two capacitors (in V), 
 $i_L$ is the current that flows in the inductor (in A), 
 $C_1$ and $C_2$ are capacitance (in F), \quad $L$ is inductance (in H), and
 $G$ is the conductance of the variable resistor (in $\Omega^{-1}$).

 The function $\tilde g(v_{C1},m_0,m_1)$ can be regarded as an active resistor. If it is locally passive, the circuit is tame, but when it is locally active, it keeps supplying power to the external circuit. The chaotic behavior is expected to be due to the power dissipated in the passive element. A difference from the van der Pol circuit is that the passive area and the active area are not separated by the strange attractor, but they are entangled.
The orbit in the $v_{C1}-v_{C2}$ plane shows a double scroll orbit or a hetero-clinic
orbit (i.e. an orbit which starts from a fixed point to another fixed point ).
 We assign $v_{C1}, v_{C2}$ and $i_L$, with a specific normalization, $x, y$ and $z$, respectively.

We did not control chaotic circuits by triggering simple electronic pulse as \cite{SaNa03}, when the orbit arrives in a certain region which we assigned '$e$' in Fig.8 of \cite{FuNi07} from which complication of the orbit starts. Main difference from \cite{SaNa03} was that we have chosen a proper height of the pulse, so that the periodic trajectory was kept and we did not need to reset the current. 
\begin{equation}
\left\{
\begin{array}{l}
\dot x=\alpha(y-h(x))\\
\dot y=x-y+z\\
\dot z=-\beta y
\end{array}
\right.
\end{equation}
where 
\[
h(x)=m_1 x+\frac{1}{2}(m_0-m_1)[|x+1|-|x-1|]
\]
\[
    h(x)=\left\{ 
      \begin{array}{cc}
        m_1x+(m_0-m_1) & x\geq 1\\
        m_0 x & |x|\leq 1\\
        m_1x-(m_0-m_1) & x\leq -1 \nonumber
    \end{array} \right.
\]
is a piecewise linear equation.

 We observed that periodic double scroll is running on heteroclinic orbits. A periodic
double scroll can be moved to a homoclinic orbit (i.e. an orbit which has single fixed point)
by triggering a pulse when the orbit arrives at a certain region. 
A simulation of the Chua's circuit was done in \cite{ZhLi97}.
Sinusoidally driven Chua's circuit was shown to have the Arnold's tongue and devil's staircase structure in the dynamical structure of the voltage\cite{PZC94}, and the sinusoidally-driven double scroll Chua circuit was analyzed by Baptista and Caldas\cite{BaCa98}. 

When a compact set $S$ is covered by $N(r)$ cubes of edge length $r$,  and each cube $i$ is associated with the probability $p_i(r)$, we define the capacity dimension $D_0$ as
\[
D_0=\lim_{r\to 0}\frac{N(r)}{|\log r|},
\]  
and the information dimension $D_1$ as
\[
D_1=\lim_{r\to 0}\frac{H(r)}{\log r}
\]
where 
\[
H(r)=\sum_{i=1}^{N(r)} p_i(r)\log p_i(r)
\]

In \cite{DGO89}, a differential equation of a pendulum
\begin{eqnarray}
&&\frac{d^2 \phi(t)}{dt^2}+\nu \frac{d\phi(t)}{dt}+g\sin\phi(t)\nonumber\\
&&=F+G\sin(\omega_1t+\alpha_1)+H\sin(\omega_2t+\alpha_2),
\end{eqnarray}
was studied, where $\omega_1$ and $\omega_2$ are incommensurate. They defined $\omega=\omega_1/\omega_2$ and considered a discrete map
\begin{eqnarray}
\phi_{n+1}&=&F(\phi_n,\theta_n)\\
\theta_{n+1}&=&[\theta_n+2\pi\omega]
\end{eqnarray}
where a square bracket indicates modulo $2\pi$. They conjectured that $D_0=2$ and $D_1=1$.  The structure of the attractor of dynamical systems whose Lyapunov exponents are non positive and which is defined by the two variables $(x,\theta)$ was studied by \cite{HeHa94,Kel96,Gle02} and \cite{AlMi08}.

The conjecture on $D_0$ and $D_1$ were confirmed in \cite{GrJa13}.
Mechanisms that changes the stable attractor to unstable attractor, which we study in a specific case of devil's staircase are important.  
Bifurcation and routes to chaos in other sinusoidally-driven extended Chua circuits was discusssed by Thamilmaran and Lakshmanan\cite{ThLa02}. 
In 2010, Leonov, Vagaitsev and Kuznetsov\cite{LVK10,LK13} claimed that there appear hidden attractors in a $n$ dimensional system with continuous vector function $\Psi(x)$ ($\Psi(0)=0$) and a constant  $n\times n$ matrix $P$,
\[
\frac{d{\bf x}}{dt}=P{\bf x}+\Psi({\bf x}).
\]
To find a periodic solution close to the harmonic oscillation, they considered a matrix $K$, such that $P_0=P+K$ has a pair of purely imaginary eigenvalues $\pm i\omega_0$ ($\omega_0>0$) and other eigenvalues of $P_0$ have negative real parts. They introduced a finite sequence of continuous functions $\phi^0({\bf x}),\phi^1({\bf x}),\cdots,\phi^m({\bf x})$, such that $\phi^0({\bf x})$ is small and $\phi^m({\bf x})=\phi({\bf x})$.They took sufficiently long period $[0,T]$ and considered
\[
\frac{d{\bf x}}{dt}=P_0{\bf x}+\phi^0({\bf x}), \qquad \frac{d{\bf x}}{dt}=P_0{\bf x}+\phi^j({\bf x}),\quad (j=1,\cdots,m)
\]
with initial condition ${\bf x}^j(0)={\bf x}^{j-1}(T)$. They showed that chaotic attractors can be produced by the perturbation. 
The review of chaotic hidden attractors in Chua's circuit is given in\cite{LK13}.

\section{Chaos in Muthuswamy-Chua's circuits}
In the study of chaotic behaviors in electronic circuits, Chua's electronic devices have attracted
constant research interests.

In 2010,  Muthuswamy and  Chua\cite{MuCh10} showed that a system with an inductor, 
capacitor and non-linear memristor can produce a chaotic circuit. The three-element
circuit with the voltage across the capacitor $x(t)=v_C(t)$, the current through the inductor
 $y(t)=i_L(t)$ and the internal state of the memristor $z(t)$ satisfy the equation
\begin{equation} \label{memeq}
\left\{
\begin{array}{l}
\dot x=\frac{y}{C}\\
\dot y=\frac{-1}{L}[x+\beta(z^2-1)y]\\
\dot z=-y-\alpha z+y\,z.
\end{array}
\right.
\end{equation}
 
We produced the similar circuit element as Muthuswamy-Chua, measured the current through the inductor, and the voltage across the capacitor.
We choose
\begin{eqnarray} 
C&=&I_s C_n T_s\nonumber\\
L&=&\frac{L_n T_s}{I_s}\nonumber\\
\beta&=&\frac{\beta_{5 kpot}}{R}\nonumber\\
\alpha&=&\frac{1}{T_s C_f \alpha_{10kpot}}
\end{eqnarray}
where $I_s=10000$, $C_n=1$nF, $T_s=10^5$, $L_n=360$mH, 
$R=1$k$\Omega$, $C_f=10$nF. 

\begin{figure}[htb]
\begin{minipage}[b]{0.47\linewidth}
\begin{center}
\includegraphics[width=6cm,angle=0,clip]{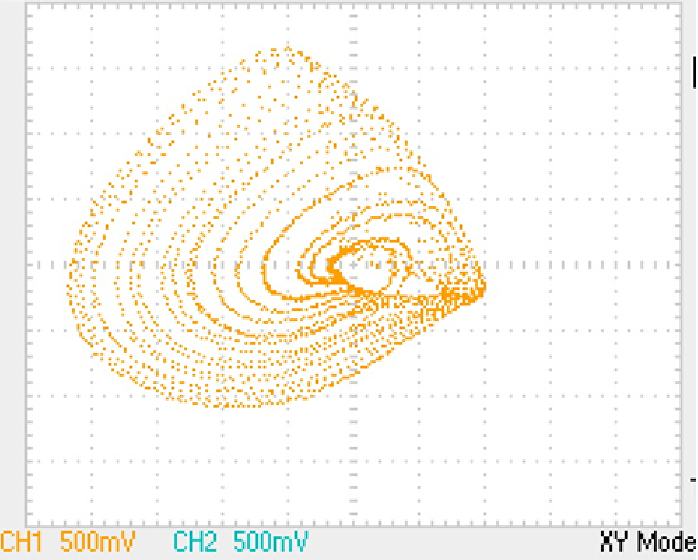}
\caption{The chaotic attractor $i_L(t)$ versus $v_C(t)$. Experimental result. .}
\label{memristor_ex}
\end{center}
\end{minipage}
\hfill
\begin{minipage}[b]{0.47\linewidth}
\begin{center}
\includegraphics[width=6cm,angle=0,clip]{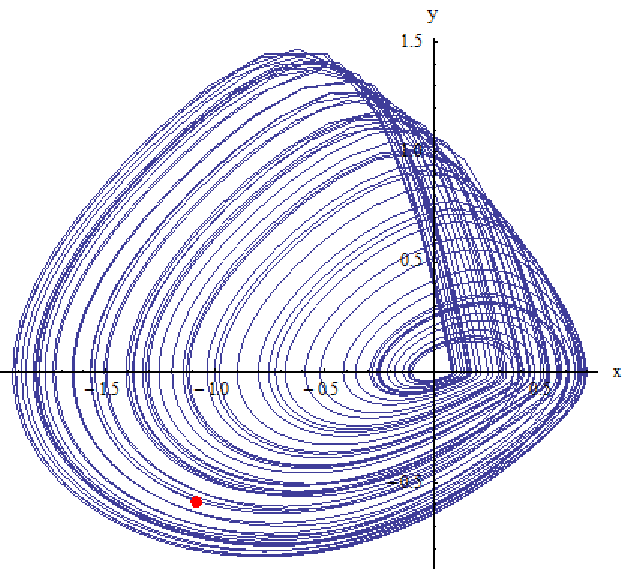}
\caption{The attractor in the xyz-space of C=1,L=3.3,$\alpha=0.2, \beta=0.5$. Numerical simulation.}
\label{Lc1L33a}
\end{center}
\end{minipage}
\end{figure}

The chaotic attractor $i_L(t)=y$ versus $v_C(t)=x$ obtained by experiment is shown in Fig.\ref{memristor_ex}. 
Here we took $\beta=1.43$

\begin{figure}[htb]
\begin{center}
\includegraphics[width=12cm,angle=0,clip]{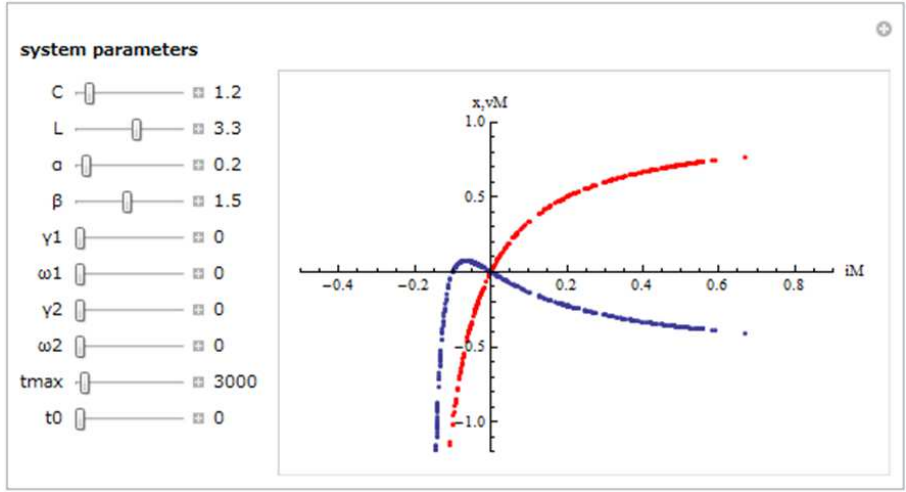} 
\caption{The DC $v_M-i_M$ and $x=iM/(iM+\alpha)$ plot at $C=1,L=3.3, \alpha=0.2$. }
\label{Lc1L33b}
\end{center}
\end{figure}
\begin{figure}[htb]
\begin{center}
\includegraphics[width=6cm,angle=0,clip]{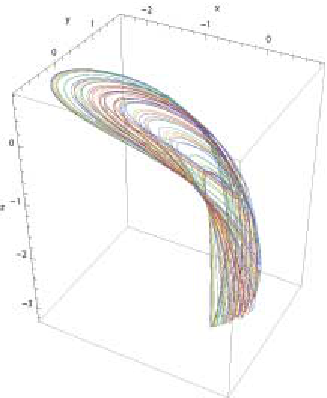} 
\caption{The attractor in the xyz-space of C=1,L=3.3,$\alpha=0.2, \beta=0.5$. Numerical simulation.}
\end{center}
\label{memristor_xyz}
\end{figure}

\begin{figure}[htb]
\begin{center}
\includegraphics[width=15cm,angle=0,clip]{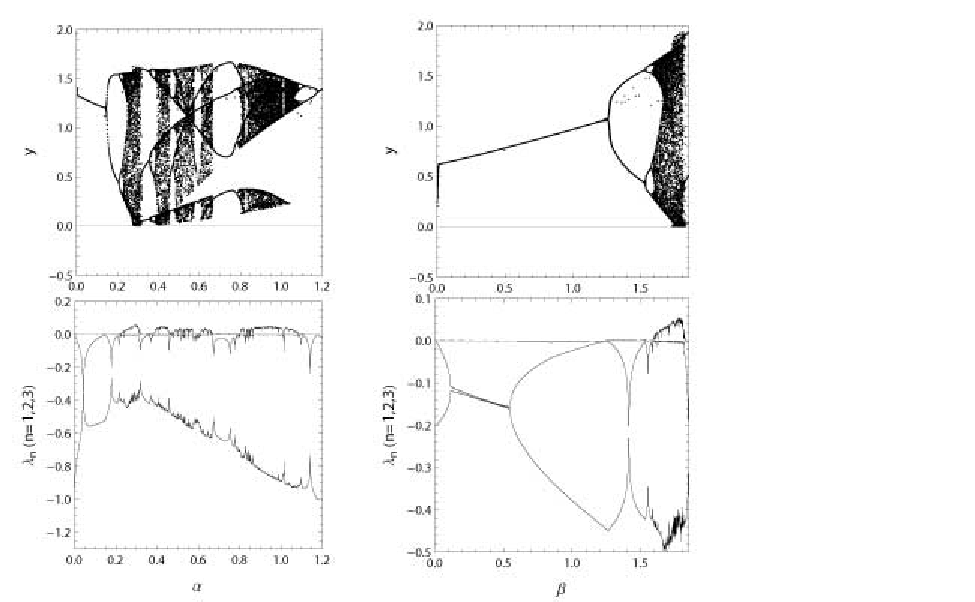} 
\caption{The bifurcation diagram and the Lyapunov exponent of Memristor. C=1,L=3.3}
\label{mem_lyap_C1}
\end{center}
\end{figure}
\begin{figure}[htb]
\begin{center}
\includegraphics[width=15cm,angle=0,clip]{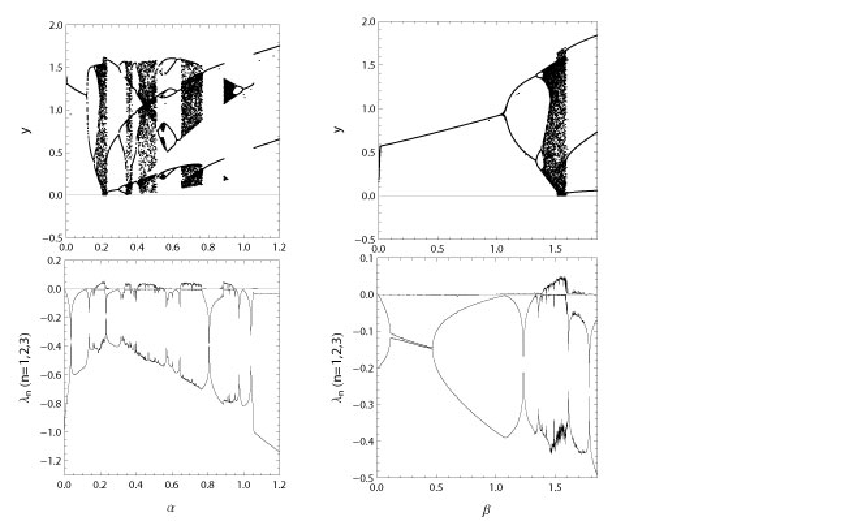} 
\caption{The bifurcation diagram and the Lyapunov exponent of Memristor. $C=1.2,L=3.3$.}
\label{mem_lyap_C2}
\end{center}
\end{figure}

When $\dot z=0$, the eq. (\ref{memeq}) becomes $-\alpha z+(z-1)y=0$ and 
$\displaystyle z=\frac{y}{y-\alpha}$.

Muthuswamy-Chua defined memrister equation by replacing $y$ by $i_M$ and $v_L+v_c=v_M$ as
\begin{eqnarray}
v_M(t)&=&\beta(z^2-1)i_M(t)\nonumber\\
\dot z&=&i_M(t)-\alpha z-i_M(t) z 
\end{eqnarray}
and by choosing $\dot z=0$, the DC characteristic becomes 
\begin{equation}\label{DCchar}
v_M(t)=\beta( \frac{i_M(t)^2}{(i_M(t)+\alpha)^2}-1)i_M(t) 
\end{equation}

When $C=1, L=3.3$ and $\alpha=0.2$, and the DC characteristic is chosen, solution of the
 eq.(\ref{DCchar}) which yields Lorenz  like attractor becomes Fig.\ref{Lc1L33a}, and the DC $v_M-i_M$ plot becomes Fig.\ref{Lc1L33b}.

The bifurcation diagram and the Lyapunov exponent $\lambda$ for $C=1, L=3.3$ and $\alpha=0.2$ as a function of $\beta$ are shown in Fig.\ref{Lc1L33b}. The diagram shows  that a bifurcation occurs at $\beta=1.23, 1.52$ and 1.58 and twice circuits, fourfolds circuits and eightfolds circuits, respectively appear.

The Fig.\ref{mem_lyap_C1} shows that in the region $1.5<\beta<1.8$ the Lyapunov
exponent $\lambda$ is positive and about 0.04, and the bifurcation diagram is chaotic.

 Ginoux, Letelier and Chua\cite{GLC10} calculated the flow curvature manifold of the
memristor. Our result with $\gamma=0$ is consistent with their results. 

\section{Strange nonchaotic attractor in driven memristive circuit}
We added external input voltage $F=\gamma \sin\omega t$ and considered the system of the equation
\begin{equation}  \label{memristor}
\left\{
\begin{array}{l}
\dot x=\frac{y}{C}\\
\dot y=\frac{-1}{L}[x+\beta(z^2-1)y+\gamma \sin\omega t]\\
\dot z=-y-\alpha z+y\,z
\end{array}
\right.
\end{equation}
 
\begin{figure}[htb]
\begin{minipage}[b]{0.47\linewidth}
\begin{center}
\includegraphics[width=8cm,angle=0,clip]{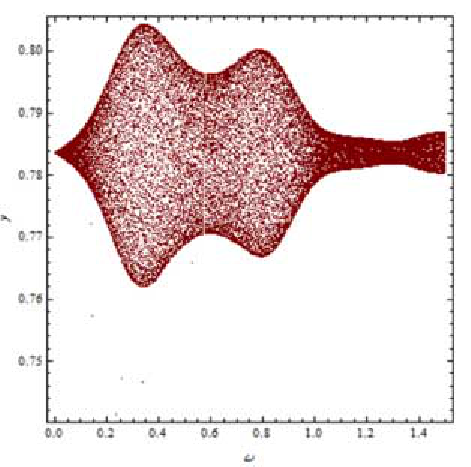} 
\caption{The bifurcation diagram $C=1,L=3.3,\alpha=0.2, \beta=0.5, \gamma=0.01$.}
\label{c1L33a02b05a}
\end{center}
\end{minipage}
\hfill
\begin{minipage}[b]{0.47\linewidth}
\begin{center}
\includegraphics[width=8cm,angle=0,clip]{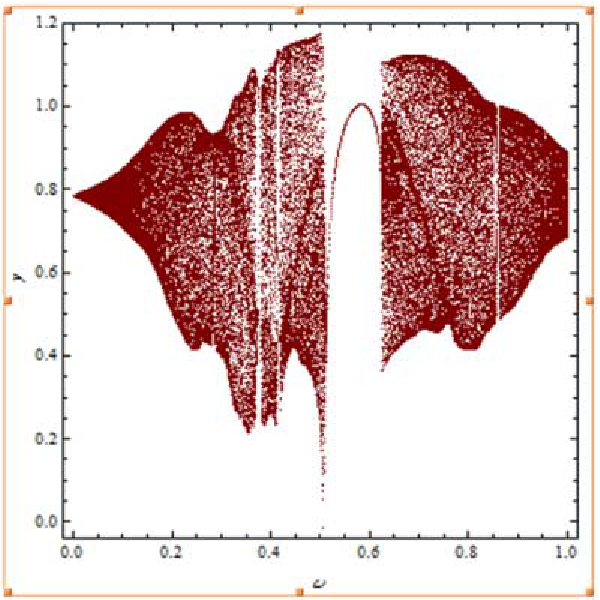} 
\caption{The bifurcation diagram $C=1,L=3.3,\alpha=0.2, \beta=0.5, \gamma=0.2$.}
\label{c1L33a02b05}
\end{center}
\end{minipage}
\end{figure}

Strange nonchaotic attractor in periodically driven systems were investigated  in \cite{PF97}
and it was argued that without external oscillation, i.e. in autonomous systems it is difficult to find a SNA, but in periodically driven systems, SNAs had been observed. We want to investigate details of chaos and SNAs in driven memristive circuits.
When $\gamma=0.01$ i.e. very small,  dependence of $y$ as a function of $\omega$ is chaotic. 
However, when we fixed $C=1.2$ and  $\gamma=0.2$, and modified $\omega$ from $0.01$ to $\omega=1$ we found a wide window of period 1 at around $\omega=0.51$, as shown in Fig.\ref{c1L33a02b05}.   

In the region $0.51\leq\omega\leq 0.62$ the period of the external voltage agrees with the 
eigenperiod of the system
\begin{equation}
\frac{1}{LC}T_s=\frac{1}{\sqrt{3.3}}T_s=0.55T_s.
\end{equation}
 The corresponding frequency is $ f\omega T_s=7.3$kHz (8.8kHz in the case of $C=1$).

In Fig.\ref{c1L33a02b05a}, we present the bifurcation diagram of the system with $C=1,L=3.3, \alpha=0.2, \beta=0.5, \gamma=0.2$. We find nonchaotic strange attractor in this case.

\begin{figure}[htb]
\begin{minipage}[b]{0.47\linewidth}
\begin{center}
\includegraphics[width=8cm,angle=0,clip]{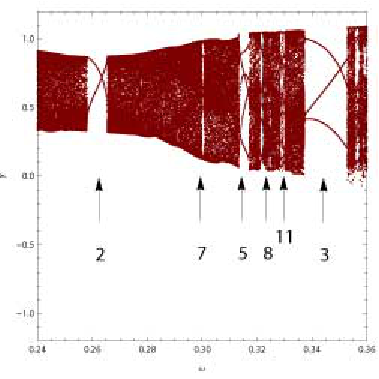}
\caption{The bifurcation diagram of the driven memristor.
$C=1.2,L=3.3, 0.24\leq \omega\leq 0.36$.}
\label{membif1}
\end{center}
\end{minipage}
\hfill
\begin{minipage}[b]{0.47\linewidth}
\begin{center}
\includegraphics[width=8cm,angle=0,clip]{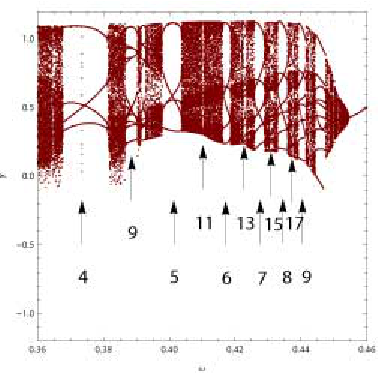}
\caption{The bifurcation diagram of the driven memristor.
$C=1.2,L=3.3,0.36\leq \omega\leq 0.46$.}
\label{membif2}
\end{center}
\end{minipage}
\end{figure}

The bifurcation diagrams of the system with $C=1.2, L=3.3, \alpha=0.2, \beta=0.5, \gamma=0.2$, in the range $0.24\leq \omega\leq 0.36$ and $0.36\leq \omega\leq 0.46$ are shown in Fig.\ref{membif1} and in Fig.\ref{membif2}, respectively, together with the frequency of the response $f_d$.

\begin{figure}[htb]
\begin{center}
\includegraphics[width=14cm,angle=0,clip]{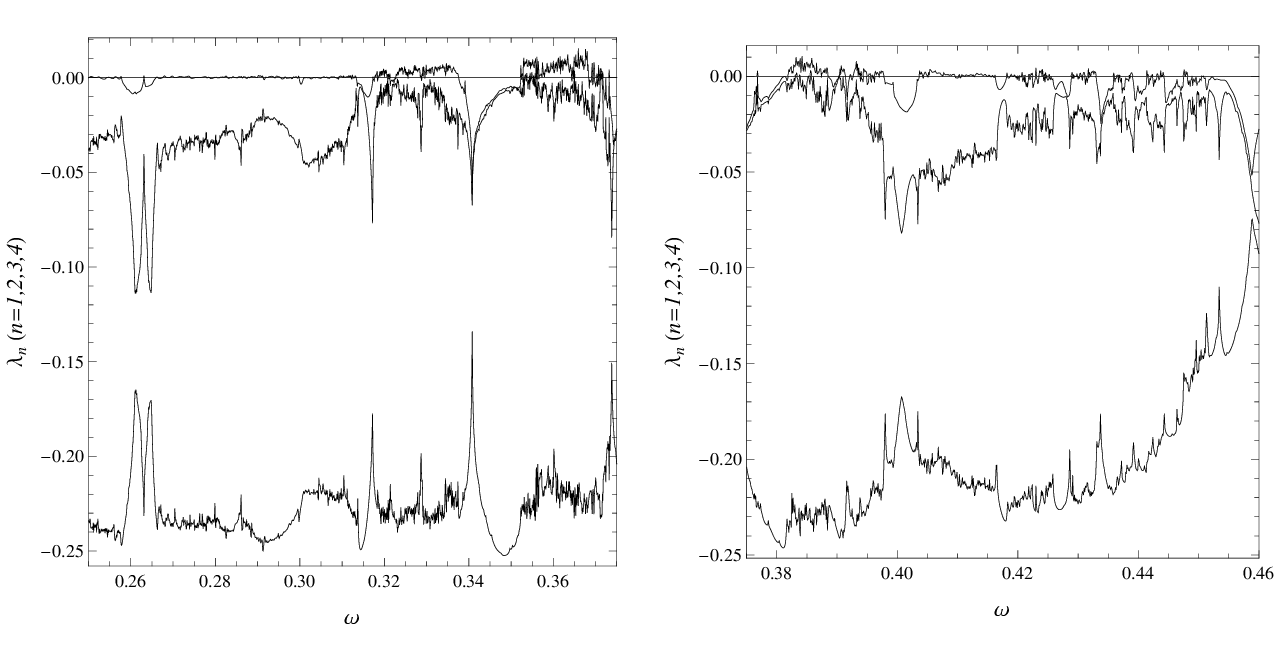}
\caption{The Lyapunov exponent of Memristor. $C=1.2,L=3.3,\alpha=0.2,\beta=0.5,\gamma=0.2$.}
\label{lyap_C12L33G02}
\end{center}
\end{figure}

We present the Lyapunov exponent $\lambda$ in Fig.\ref{lyap_C12L33G02}.  When $\omega\sim 0.3605$, $\lambda$ becomes slightly positive up to about 0.01. The Lyapunov exponents are negative in other windows.

\begin{figure}[htb]
\begin{minipage}[b]{0.47\linewidth}
\begin{center}
\includegraphics[width=8cm,angle=0,clip]{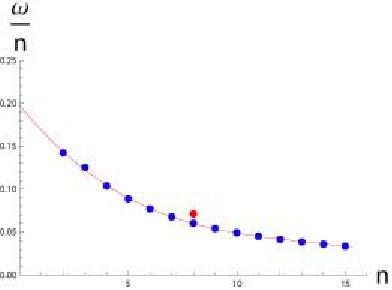}
\caption{The fit of $\omega/n$ as a function of the number of nodes $n$. $C=1$}
\label{omg_n1}
\end{center}
\end{minipage}
\hfill
\begin{minipage}[b]{0.47\linewidth}
\begin{center}
\includegraphics[width=8cm,angle=0,clip]{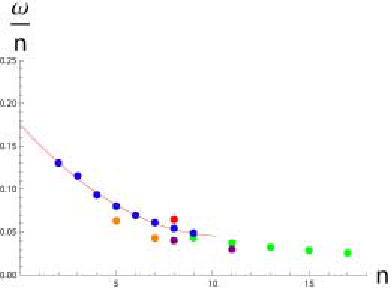} 
\caption{The fit of $\omega/n$ as a function of number of nodes $n$. $C=1.2$}
\label{omg_n2}
\end{center}
\end{minipage}
\end{figure}

The frequency of the windows $\omega$ as a function of the number of periodicity of the response for samples of  $C=1$ is shown in Fig.\ref{omg_n1}. In the case of one period, we put the $\omega$ at $f_d=8$, since when we put $\omega/f_d$ as a function of the number of periodicity $f_d$, the $\omega$ of one periodicity divided by 8 becomes close to that of periodicity 7. The $\omega/n$ for samples of $C=1.2$ are shown in Fig.\ref{omg_n2}.  It contains the series from $f_d=2$ to 9, from 9 to 17,  from 5 to 7 and from 8 to 11, which are shown by different colors.
 The single period window may be considered as 8 degenerate nodes. The energy  $\omega/8$ of this node put at $n=8$ is consistent with $\omega/f_d$ of $f_d=7$, i.e. energy per node of  $f_d=Mod(-1,8)=7$. 

The matching $\omega/f_d$ in the case of $C=1$ is about 10\% lower than that of $C=1.2$, and the width of the window of $C=1$ is narrower than that of $C=1.2$.

\begin{figure}[htb]
\begin{minipage}[b]{0.47\linewidth}
\begin{center}
\includegraphics[width=8cm,angle=0,clip]{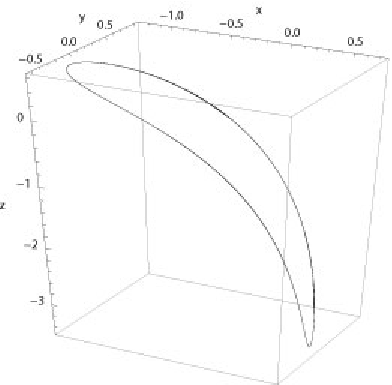} 
\caption{The attractor of $n=1$ in the $(x(t),y(t),z(t))$ space. $f_d=1$. $C=1.2,L=3.3,\alpha=0.2,\beta=0.5, \gamma=0.2$}
\label{xyz_51}
\end{center}
\end{minipage}
\hfill
\begin{minipage}[b]{0.47\linewidth}
\begin{center}
\includegraphics[width=8cm,angle=0,clip]{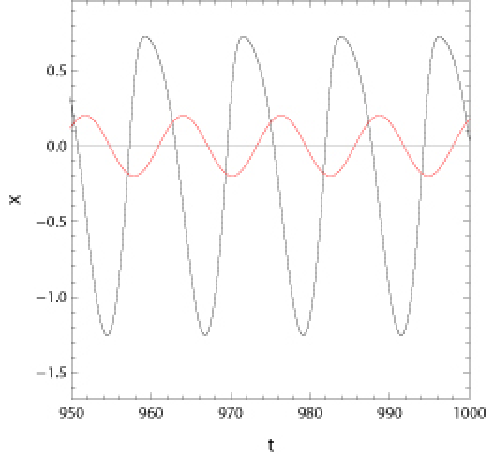} 
\caption{The time series of $x(t)$(black) and $\gamma\sin\omega t$(red). $\omega=0.51, W=1/1, C=1.2,L=3.3$}
\label{xtimes_n1}
\end{center}
\end{minipage}
\end{figure}

The attractor of $n=1$ of samples $C=1.2,L=3.3$ at $\omega=0.51$ is shown in Fig.\ref{xyz_51}, and the time series of $x(t)$ is compared with $\gamma\sin\omega t$ in Fig.\ref{xtimes_n1}. 

Since the frequency of the response and that of driving term are both equal to 1, the ratio of $W=f_s/f_d=1/1$

At windows of nodes of $f_d=2$ to $f_d=9$ of samples $C=1.2,L=3.3$, we compared the time series $x(t)$ of the response and input $\gamma\sin\omega t$. The observed ratio $W=f_s/f_d$ turned out to be
\[
\{\frac{1}{2}, \frac{2}{3}, \frac{3}{4}, \frac{4}{5},  \frac{5}{6}, \frac{6}{7}, \frac{7}{8}, \frac{8}{9},\frac{1}{1}\}.
\]
The sequence follows the period adding law\cite{PZC94}
\[
\frac{q}{p}\to \frac{q+Q}{p+P}\to \frac{q+2Q}{p+2P}\to\cdots\to {\rm chaos}\to \frac{Q}{P}.
\]
 
The time series $x(t)$ and $\gamma\sin\omega t$ at $\displaystyle \omega={0.4}$ allows an assignment of  $W=4/5$.

Between the window of $f_d=2$ and $f_d=3$, we found a window of $f_d=5=2+3$, which appears from the Farey sum 
\[
\frac{1}{2} + \frac{2}{3}\to \frac{3}{5}
\]
which satisfys the relation $\displaystyle \frac{1}{2}<\frac{3}{5}<\frac{2}{3}$.
In general when $\displaystyle \frac{a}{b}<\frac{c}{d}$ and $bp-qa=qc-dp=1$,
$bp+dp=qc+qa$ or $(b+d)p=(c+a)q$ and $\displaystyle\frac{a+c}{b+d}=\frac{p}{q}$
and $\displaystyle \frac{a}{b}<\frac{p}{q}<\frac{c}{d}$.

\begin{figure}[htb]
\begin{minipage}[b]{0.47\linewidth}
\begin{center}
\includegraphics[width=8cm,angle=0,clip]{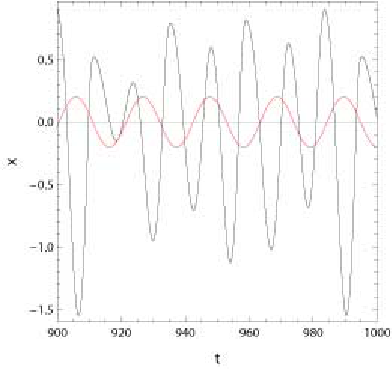} 
\caption{The time series of $x(t)$)black) and $\gamma\sin\omega t$(red). $\omega=0.30, W=4/7, C=1.2,L=3.3$}
\label{xtimes_n74}
\end{center}
\end{minipage}
\hfill
\begin{minipage}[b]{0.47\linewidth}
\begin{center}
\includegraphics[width=8cm,angle=0,clip]{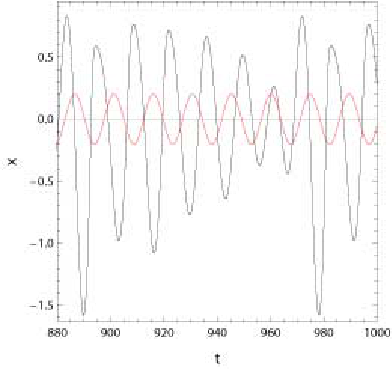} 
\caption{The time series of $x(t)$(black) and $\gamma\sin\omega t$(red). $\omega=0.427, W=6/7, C=1.2,L=3.3$}
\label{xtimes_n76}
\end{center}
\end{minipage}
\end{figure}

A comparison of the time series $x(t)$ and the corresponding $\gamma\sin\omega t$ at $\displaystyle \omega=0.315$  allows the assignment $W=3/5$. 
This new node of $W=3/5$ and the node of $W=1/2$ make a node of $f_d=7=5+2$.
A comparison of the time series $x(t)$ and the corresponding $\gamma\sin\omega t$ at $\displaystyle\omega={0.300}$ allows the assignment $W=4/7$ as shown in Fig.\ref{xtimes_n74}.  The new $W$ again appears from the Farey sum
\[
 \frac{1}{2} + \frac{3}{5}\to \frac{4}{7} 
\]
The time series of $x(t)$ and $\gamma\sin\omega t$ of $W=6/7$ are shown in Fig.\ref{xtimes_n76} for comparison.

\begin{figure}[htb]
\begin{minipage}[b]{0.47\linewidth}
\begin{center}
\includegraphics[width=8cm,angle=0,clip]{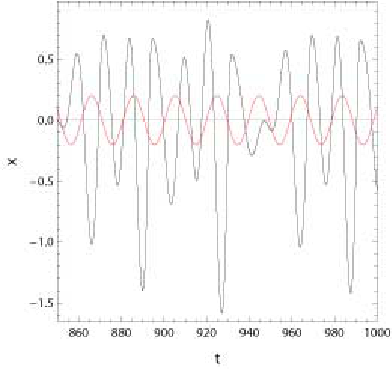} 
\caption{The time series of $x(t)$(black) and $\gamma\sin\omega t$(red). $\omega=0.321, W=5/8, C=1.2,L=3.3$}
\label{xtimes_n85}
\end{center}
\end{minipage}
\hfill
\begin{minipage}[b]{0.47\linewidth}
\begin{center}
\includegraphics[width=8cm,angle=0,clip]{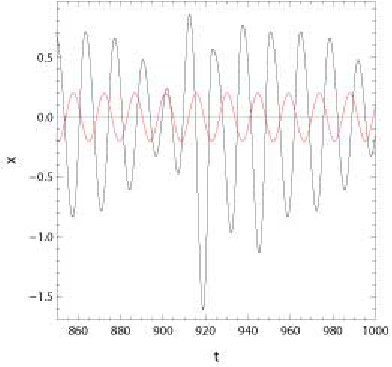} 
\caption{The time series of $x(t)$(black) and $\gamma\sin\omega t$(red). $\omega=0.434, W=7/8, C=1.2,L=3.3$}
\label{xtimes_n87}
\end{center}
\end{minipage}
\end{figure}

A similar comparison of  $f_d=8$ data, $\displaystyle W=\frac{2}{3}+\frac{3}{5}\to\frac{5}{8}$ and $\displaystyle W=\frac{7}{8}$ are shown in Fig.\ref{xtimes_n85} and Fig.\ref{xtimes_n87}, respectively.

We compared, also $f_d=9$ data, $\displaystyle W=\frac{3}{4}+\frac{4}{5}\to\frac{7}{9}$ and $\displaystyle W=\frac{8}{9}$.

 The series 
\[
\{\frac{3}{5},\frac{4}{7},\cdots,\frac{1}{2}\} 
\]
follows the period adding law\cite{PZC94} with $q=3,Q=1,p=5,P=2$.
The series
\[
\{\frac{5}{8},\frac{7}{9},\cdots,\frac{2}{1}\}
\]
follows the period adding law with $q=5,Q=2,p=8,P=1$.

There are other Farey sums
\[
\frac{4}{5}+\frac{5}{6}\to\frac{9}{11}, \quad{\rm upto}\quad \frac{7}{8}+\frac{8}{9}\to \frac{15}{17}
\]
from $\omega=0.41$ until $\omega=0.4375$, which make the series of $W$
\[
\{\frac{9}{11}, \frac{11}{13}, \frac{13}{15}, \frac{15}{17},\frac{2}{2}\}.
\]
Near the region of $\omega=0.45$, there is a window which has 10 nodes from left cluster
and 9 nodes from right cluster as shown in Fig.\ref{membif2}. We assign $\displaystyle W=\{\frac{9}{9},\frac{8}{8},\frac{7}{7}\}$ at $\omega=0.446,0.450$ and 0.4515, respectively, and the Farey sums
\[
\frac{8}{9}+\frac{9}{9}\to\frac{17}{18}, \quad{\rm and}\quad \frac{9}{9}+\frac{8}{8}\to\frac{17}{17}
\]
at $\omega=0.4425$ and $\omega=0.448$, respectively. 

The bifurcation diagram shows that when $\omega=0.3605$, the system is chaotic.
We studied Poincar\'e section of driven memristive circuit at this  $\omega=0.3605$ by taking the Poincar\'e section on the $xy$ plane, the $yz$ plane and the $xz$ plane.

\begin{figure}[htb]
\begin{minipage}[b]{0.47\linewidth}
\begin{center}
\includegraphics[width=8cm,angle=0,clip]{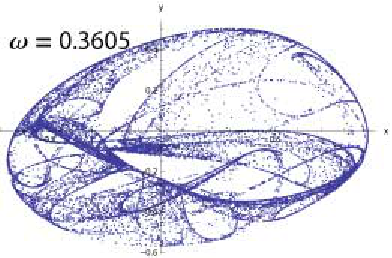} 
\caption{The Poincar\'e section on the $xy$ plane. $\omega=0.3605$.}
\label{poincare_g1}
\end{center}
\end{minipage}
\hfill
\begin{minipage}[b]{0.47\linewidth}
\begin{center}
\includegraphics[width=8cm,angle=0,clip]{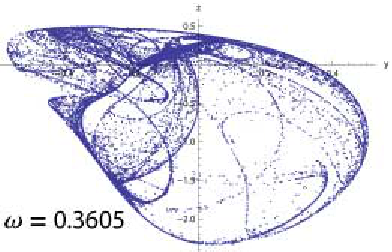} 
\caption{The Poincar\'e section on the $yz$ plane.  $\omega=0.3605$.}
\label{poincare_g2}
\end{center}
\end{minipage}
\end{figure}

\begin{figure}[htb]
\begin{center}
\includegraphics[width=8cm,angle=0,clip]{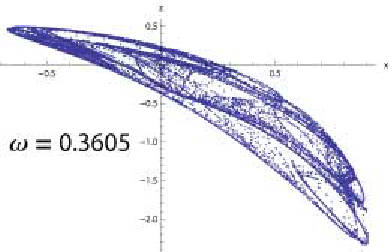} 
\caption{ThePoincar\'e section on the $xz$ plane. $\omega=0.3605$.}
\label{poincare_g3}
\end{center}
\end{figure}

The points are plotted, when $\omega t$ satisfies $Mod[\omega t, 2\pi]=0$. 
We calculated the Poincar\'e section of $\displaystyle \omega=0.3605$ on the $xy$, the $yz$ and the $xz$ plane, which are shown in Fig.\ref{poincare_g1}, \ref{poincare_g2} and \ref{poincare_g3}, respectively. 

In the case of oscillation on the $xy$ plane, we calculated $\Theta_n=\tan^{-1}(y_n/x_n)$, where $x_n$ and $y_n$ are the $x$ and $y$ coordinates of the Poincar\'e map. 
The return map of $\Theta_{n+1}$ v.s. $\Theta_n$ simplifies the chaotic behaviors. 
  It becomes a single line in the case of $\omega<0.26$,  four clusters when $\omega=0.3$, and chaotic when $\omega={0.32}$. 
 When $\displaystyle \omega=0.5$, clusters appear again, and when $\displaystyle \omega=0.6$, a single line appears again. 

We observed that except near $\displaystyle \omega={0.36}$,  $\lambda$ is negative, and the oscillation is non-chaotic in most regions.


In the case of $C=1.2$, we observed windows=$\{ \displaystyle \frac{1}{2},\cdots,\frac{8}{9}\}$,i.e. 8 levels were observed, (The level  $\displaystyle \frac{9}{10}$ is broken in the high-frequency part.) and above the level $\displaystyle \frac{4}{5}$ the level $\displaystyle \frac{7}{9}$ in the Farey sequence $\displaystyle\{\frac{3}{4},\frac{7}{9},\frac{4}{5}\}$ is made and the sequence
$\{\displaystyle \frac{9}{11},\cdots,\frac{15}{17}\}$ follows.

Below the level $\displaystyle\frac{7}{9}$ the Farey sequences are different, and as the source of 8 levels in the $C=1.2$ case, we imagine a combination of the input electron and the output electron both expressed by quaternions. Two quaternions can make an octonion which couples with 8 dimensional vector fields. 

In the case of $C=1.0$ and $\gamma=0.2$, $\omega$ of the node $\displaystyle W=\frac{1}{1}$ is about 10\% larger than that of $C=1.2$, and the node $n$ at highest $\omega$ was assigned up to 15, in contrast to up to 8 in the case of $C=1.2$. 

The width of the windows is narrower, and the sequences of $W$ consist of
\[
\{\frac{1}{2},\frac{2}{3},\frac{3}{4},\frac{4}{5},\frac{5}{6},\frac{6}{7},\frac{7}{8},\frac{8}{9},\frac{9}{10},\frac{10}{11},\frac{11}{12},\frac{12}{13},\frac{13}{14},\frac{14}{15},\frac{1}{1}\}.
\]
The Farey sequences of  $P\ne Q$ like $\displaystyle W=\{\frac{3}{5},\frac{4}{7}\}$ and $\displaystyle W=\{\frac{5}{8},\frac{7}{9}\}$, which were observed in $C=1.2$, were not observed in $C=1.0$.

\begin{figure}[htb]
\begin{minipage}[htb]{0.47\linewidth}
\begin{center}
\includegraphics[width=8cm,angle=0,clip]{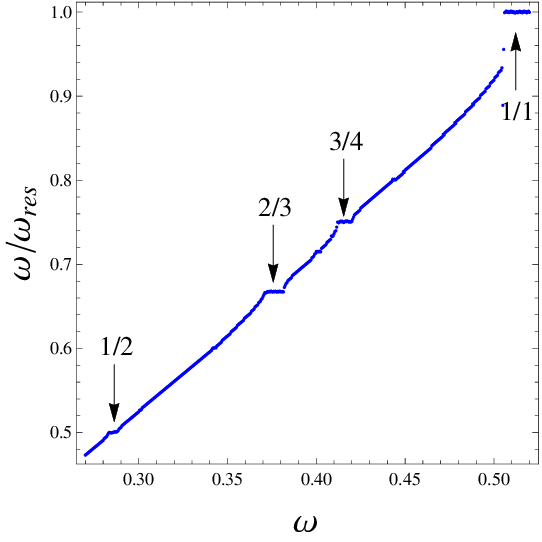}
\caption{The devil's staircase for $C=1$.} 
\label{d1}
\end{center}
\end{minipage}
\hfill
\begin{minipage}[htb]{0.47\linewidth}
\begin{center}
\includegraphics[width=8cm,angle=0,clip]{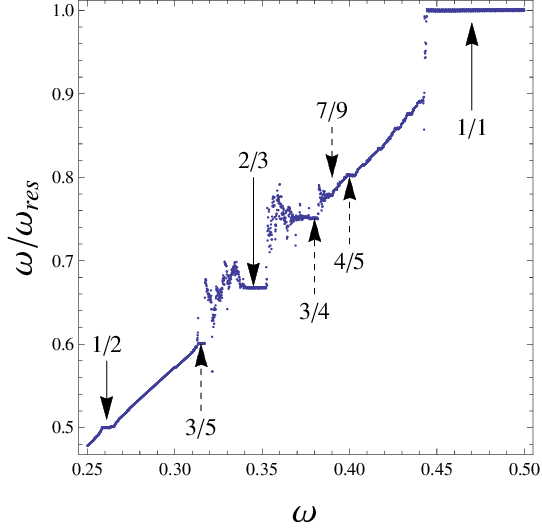}
\caption{The devil's staircase for $C=1.2$.}
\label{d2}
\end{center}
\end{minipage}
\end{figure}

Extending this analysis to the lower frequency region, we observed the devil's staircase structure\cite{Santos98} in the case of $C=1$ as Fig.\ref{d1} and in the case of $C=1.2$, as Fig.\ref{d2}. 

The qualitative difference of the case of $C=1$ and $C=1.2$ can be explained by the hidden attractor\cite{LVK10,LK13}. They showed that when there is a perturbation with amplitude small enough, the Poincar\'e map of  $F$ of the set $\Omega$ of solutions of the differential equation into itself
\[
F\Omega\subset \Omega
\]
can introduce stable solutions. When $C=1.2$ the $F\Omega$ goes outside $\Omega$ and unstable solutions dominate, but when $C=1$, there appear solutions of $f_s<f_d$ and hidden attractors appear.
As shown in \cite{HeHa94}, the birth of SNA is due to collision of an unstable parent torus and a period-doubled torus. In our system, the driving oscillation $\gamma \sin\omega t$ of various $\omega$ plays the role of period-doubling torus which has, in the region $0.265<\omega<0.30$,  the Lyapunov exponent $\lambda_1=0$, $\lambda_2\sim 0$ ,$\lambda_4<\lambda_3<0$. Since the SNA is defined from the insensitivity to the initial condition, and negative Lyapunov exponents, the attractors we observed in the low frequency region which have negative Lyapunov exponents are SNA. The qualitative difference of $C=1.0$ and $C=1.2$ can be also explained by the Lyapunov exponent $\lambda_3$ of $C=1.2$ (Fig.\ref{mem_lyap_C2}) which is larger than that of $C=1.0$(Fig.\ref{mem_lyap_C1}) and the system of $C=1.2$ is more unstable than that of $C=1.0$.

\section{Conjecture on the frequency of the windows of driven Muthuswamy-Chua's circuit} 
We observed that the frequency $\omega$ of the single periodicity devided by 8 is close to the frequency $\omega$ of the seven periodicity devided by 7 in Fig.\ref{omg_n1} and \ref{omg_n2}. It suggests that the single periodicity is equivalent to 8 periodicity and the octonion plays an important role in the interaction of electrons which can be expressed by Dirac spinors and electromagnetic fields.

The four component Dirac's spinor is a combination of two two-component spinors each transforms as a quaternion.
Cartan\cite{Cartan66} defined, using semi-spinors of an even number of indices
\[
   \xi_{even}:= \xi_{12},\xi_{23}, \xi_{34}, \xi_{13}, \xi_{24}, \xi_{14}, \xi_{1234}, \xi_0
\]
and semi-spinors of an odd number of indices
\[
   \xi_{odd}:= \xi_{1}, \xi_{2}, \xi_{3}, \xi_{4}, \xi_{234}, \xi_{134}, \xi_{124}, \xi_{123},
\]
for bases of spinors

In octonion basis, we consider a fermion $\phi$ and its charge conjugate $\mC\phi$\cite{SF12,SF13a}, which are described by
\begin{eqnarray}
{\phi}&=&\xi_0 I+\xi_{14}\sigma_x +\xi_{24}\sigma_y+\xi_{34}\sigma_z\nonumber\\
&=&\left(\begin{array}{cc}A_4+i \, A_3&i \, A_1-A_2\\
i\,A_1+A_2& A_4-i\,A_3\end{array}\right)=(\vec A,A_4)\nonumber\\
{\mathcal C}\phi&=&\xi_{1234} I-\xi_{23}\sigma_x-\xi_{31}\sigma_y-\xi_{12}\sigma_z\nonumber\\
&=&\left(\begin{array}{cc} B_4-i\,B_3&-i\, B_1+B_2\\
-i\,B_1-B_2 & B_4+i\, B_3\end{array}\right)=(\vec B,B_4)\nonumber
\end{eqnarray}
and a fermion $\psi$ and its charge conjugate $C\psi$, which are described by
\begin{eqnarray}
{\psi}&=&\xi_4 I+\xi_{1}\sigma_x+\xi_{2}\sigma_y+\xi_{3}\sigma_z\nonumber\\
&=&\left(\begin{array}{cc}C_4+i \,C_{3}&i \,C_{1}-C_{2}\\
i\, C_1+C_2&C_4-i\,C_{3}\end{array}\right)=(\vec C,C_4)\nonumber\\
{\mathcal C}\psi&=&\xi_{123}I-\xi_{234}\sigma_x-\xi_{314}\sigma_y-\xi_{124}\sigma_z\nonumber\\
&=&\left(\begin{array}{cc}D_4-i \,D_3&-i \,D_1+D_2\\
-i\,D_1-D_2& D_4+i \,D_3\end{array}\right)=(\vec D,D_4)\nonumber.
\end{eqnarray}
The choice of particle v.s. hole for $\psi$ v.s. $\phi$ depends on convention. Following the usual convention, we choose $\psi$ for a particle and $\phi$ for a hole.

The coupling of the spinor and vectors defined by Cartan is,
\begin{eqnarray}
{\mathcal F}&=&X{\mathcal C}\phi\psi+X{\mathcal C}\phi{\mathcal C}\psi\nonumber\\
&=&x^1(-A_1 D_4-B_2D_3+B_3 D_2+B_4 C_1)\nonumber\\
&+&x^2(B_1 D_3-A_2 D_4-B_3 D_1+B_4 C_2)\nonumber\\
&+&x^3(-B_1 D_2+B_2 D_1-A_3 D_4+B_4 C_3)\nonumber\\
&+&x^4(-A_1 D_1-A_2 D_2-A_3 D_3+B_4 C_4)\nonumber\\
&+&x^{1'}(B_1 C_4-A_2 C_3-A_3 C_2-A_4 D_1)\nonumber\\
&+&x^{2'}(A_1 C_3+B_2 C_4-A_3 C_1-A_4 D_2)\nonumber\\
&+&x^{3'}(-A_1 C_2+A_2 C_1+B_3 C_4-A_4 D_3)\nonumber\\
&+&x^{4'}(-B_1 C_1-B_2 C_2-B_3 C_3+A_4 D_4).\nonumber
\end{eqnarray}

In our system, there are an input electron which is represented by the plane wave, and an output fermions/holes.
There are 4 possible choices of the input fermion wave functions $A,B,C,D$, and 2 possible types of exchanged photons i.e.  left vertex and right vertex are the same $x_4$ or $x_4'$ exchange. The coulomb interaction in coordinate space is instantaneous and expressed as
\[
D^C_{00}(x)=\frac{1}{4\pi}\frac{1}{|\vec x|}\delta(x_0)
\]
and it is natural to indistinguish $x_4$ and $x_4'$ and allow left vertex is that of $x_4$ and right vertex is that of $x_4'$ or vice versa, which we denote $x_4/x_4'$ exchange.

\begin{figure}[htb]
\begin{minipage}[b]{0.47\linewidth}
\begin{center}
\includegraphics[width=4cm,angle=0,clip]{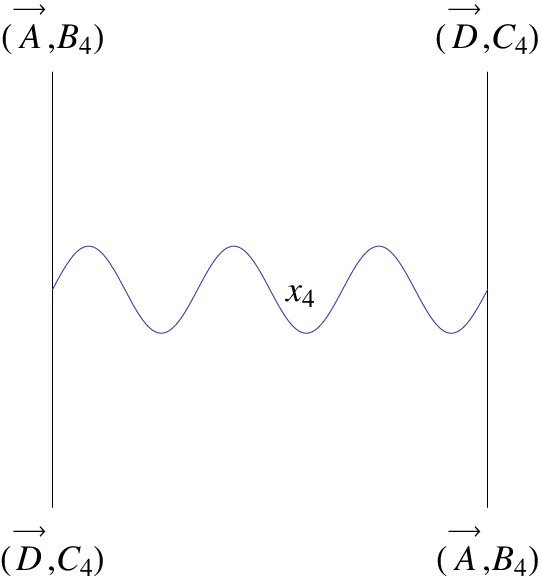}
\caption{The vector $x_4$ exchange between a partcle-particle $ (\vec D,C_4),(\vec A,B_4)$.}
\label{vectorex2}
\end{center}
\end{minipage}
\hfill
\begin{minipage}[b]{0.47\linewidth}
\begin{center}
\includegraphics[width=4cm,angle=0,clip]{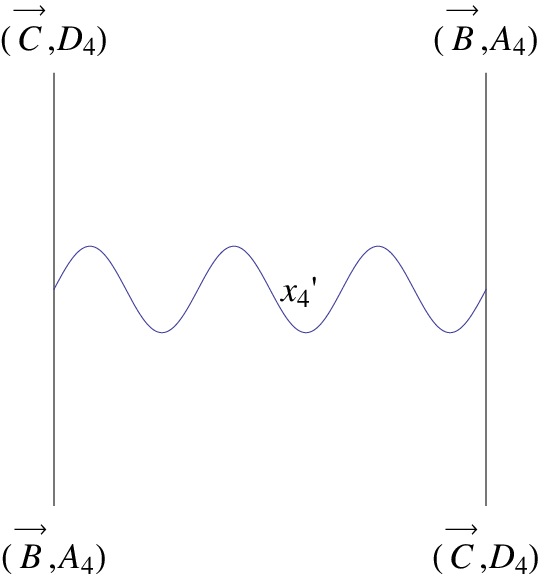}
\caption{The vector $x_4'$ exchange between a hole-hole $(\vec B,A_4),(\vec C,D_4)$.} 
\label{vectorex1}
\end{center}
\end{minipage}
\end{figure}

The exchange of $x_4$ between a hole with its 4th component interchanged by that of charge conjugate particle
\[
\tilde{\phi}=\left(\begin{array}{cc}B_4+i \, A_3&i \, A_1-A_2\\
i\,A_1+A_2& B_4-i\,A_3\end{array}\right)=(\vec A, B_4)
\] 
and another hole with its 4th component interchanged by that of charge conjugate particle
\[
\widetilde{\mathcal C \psi}=\left(\begin{array}{cc}C_4+i \, D_3&i \, D_1-D_2\\
i\,D_1+D_2& C_4-i\,D_3\end{array}\right)=(\vec D,C_4)
\]
 is represented in Fig.\ref{vectorex2}.

 The exchange of $x_4'$ between a particle with its 4th component interchanged by that of charge conjugate hole
\[
\widetilde{\mathcal C \phi }=\left(\begin{array}{cc}A_4+i \, B_3&i \, B_1-B_2\\
i\,B_1+B_2& A_4-i\,B_3\end{array}\right)=(\vec B, A_4)
\]
and another particle with its 4th component interchanged by that of charge conjugate hole
\[
\tilde{ \psi}=\left(\begin{array}{cc}D_4+i \, C_3&i \, C_1-C_2\\
i\,C_1+C_2& D_4-i\,C_3\end{array}\right)=(\vec C,D_4)
\] 
is represented in Fig.\ref{vectorex1}.
 
\begin{figure}[htb]
\begin{center}
\includegraphics[width=4cm,angle=0,clip]{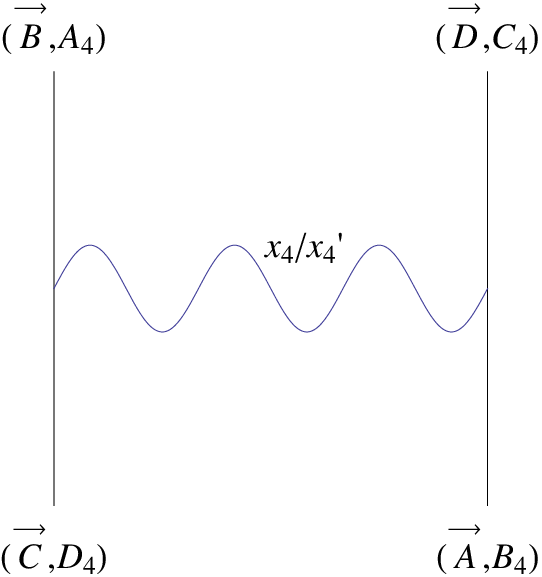}
\caption{The vector $x_4/x_4'$ exchange between a hole-particle $ (\vec C,D_4),(\vec A,B_4) $}
\label{vectorex3}
\end{center}
\end{figure}

Since the vector particle is assumed to be self-dual, or propagation of $x_4$ is same as that of $x_4'$, the hole-particle interaction between $(\vec C,D_4)$ and $(\vec A,B_4)$ represented in Fig.\ref{vectorex3}. The particle-hole interaction between $(\vec D,C_4)$ and $(\vec B,A_4)$ can be obtained by performing the time reversal.

In our analysis of memristive circuit, we defined the voltage across
the capacitor $x_1=v_C(t)$, the current through the inductor $x_2=i_L(t)$ and
the internal state of the memristor $x_3=z(t)$.

From the frequency $\omega$ of the single period oscillation observed in the widest window of the bifurcation diagram, we calculate $\omega/8$ and compare the $\omega/n$ of oscillation of periodicity equals to $n$.   
  We observed a three periodic oscillation around $\omega_3=0.375$, $\omega_3/3=0.125$, a four periodic oscillation around $\omega_4=0.416$, $\omega_4/4=0.104$, and a five periodic oscillation around $\omega_5=0.444$, $\omega_5/5=0.088$.
$\omega$ divided by a number of periods $n$ defined as $\omega/n$ is nearly linear function of $0.18-0.018 n$, and $n$ runs from 2 to 15.

In the case of $C=1.2$, we observe windows of periodicity $n$ from 2 to 9. The last window of periodicity 10 is broken at the high-frequency region. In the case of $C=1.0$, we observed windows of $n=2,\cdots,9$, which have the smooth n dependence of $\omega/n$.

\section{Discussion}
We studied the sinusoidally driven memristor circuit. We observed nonchaotic strange attractors and
 different from the case of driven Chua's circuit\cite{PZC94}, we first observed period adding low as in LCR circuit\cite{MTCP05}.
In the case of $C=1.2$, we observed a sequence of response frequency $f_d$ stepwise increasing, and the winding number $W=f_s/f_d$ shows a sequence $\displaystyle W=\{\frac{1}{2}, \frac{2}{3}, \frac{3}{4},\cdots, \frac{8}{9}, \frac{1}{1}\}$ and Farey sum sequences $\displaystyle W=\{\frac{3}{5}, \frac{4}{7},\cdots,\frac{1}{2}\}$ and $\displaystyle \{\frac{5}{8}, \frac{7}{9},\cdots, \frac{2}{1}\}$. 
It is interesting that in the analysis of non-linear circuit of \cite{CYY86}, the number of response $P$ of the $v_C$, which corresponds to our $f_d$, was assigned as Level-2 devil's staircase sequence from step 2, and 8 steps $P=2,3,\cdots,9$ were considered as in our case of $C=1.2$. The driving oscillation is $P=1$, in their cases and their responses are not hidden oscillations.

In the case of $C=1.0$, we find winding number's sequences $\displaystyle W=\{\frac{1}{2},\frac{2}{3},\frac{3}{4},\cdots, \frac{14}{15}, \frac{1}{1}\}$, but we do not find the Farey sum sequences observed in $C=1.2$. It means that the way to chaos via the devil's staircase which is observed in the case of $C=1.2$ is absent in the case of $C=1$.
In the Level-2 devil's staircase sequence from step 3 to step 2 of \cite{CYY86}, 7 steps $P=3,5,\cdots,15$ were considered and devil's staircase route to chaos was discusseed.

When $\omega$ of $W=1$ is divided by 8, it becomes close to $\omega$ of $\displaystyle W=\frac{6}{7}$ divided by 7. It suggests that the octonion that appears as a combination of two quaternions plays an important role in the circuit, since Dirac electrons are expressed by octonions.

 Oscillation in the voltage of two capacitances $v_{C1}, v_{C2}$ and the current $i_L$ of an inductor in nonlinear coupled circuits can be analyzed using the quaternion bases.   
 In our memristive circuits,  octonion plays an important role in the matching of period $\omega/f_d$. 
 Symmetry of electromagnetic field and that of probes must be studied together.
 We found nonchaotic strange attractors, or strange attractors whose Lyapunov exponent is negative, in the driven memristive circuit. 

The Poincar\'e map of the memristor current obtained by choosing the plane defined by the condition $Mod[\omega t,2\pi]=0$ shows a relatively complicated structure at a certain frequency region. When the amplitude of the external oscillation is small, the strange attractor remains nonchaotic, but when it is large, the strange attractor at a certain frequency region becomes chaotic. The condition depends upon the capacity of the condenser $C$. 

In the analysis of bifurcation of ordinary differential equation, treatment of structurally unstable invariant sets in a suitable phase-parameter space was discussed\cite{KKLN93}, and MATLAB programs using the coordinates that show collocation at Gaussian points\cite{DBS78} for bifurcation analysis of dynamical systems were proposed\cite{GKD00,DGK03,DGKMS07}. We did not choose coordinates that show collocation at Gaussian points, but since we have chosen fine mesh points in the program of Mathematica, we do not expect qualitative changes appear by different choice of mesh points.  

We considered vector particle interactions with electron particles and holes which are represented by octonions. It is not evident that octonions, give better results than a combination of quaternions, but it is worth studying the possibility of representing quarks and anti-quarks and lepton anti-leptons by octonions\cite{SF12,SF13a,SF13b,SF14a,SF15}. 

The recent analysis of a boson described by lepton and antilepton decay into two photons \cite{SF14b} shows that the description of spinors of leptons by octonions and description of interaction between vector fields in $\Vec R^3$ and spinors of Cartan is promising. 

The Coulomb interaction $D^C_{00}(x)$ appears from $\bar\psi\gamma_0\psi$ and the wave function $\psi$ is an element in the twistor space. Identity of the electroweak gauge group of Weinberg-Salam model and the gauge group of Dirac current in which $\psi$ is treated as a Clifford algebra spinor 
\[
\psi=\Psi\frac{1}{2}(1+\gamma_0)\frac{1}{2}(1+i\gamma_2\gamma_1),
\]
where $\Psi$ in ${\mathcal Cl}_{1,3}^+$ is defined as
\[
\Psi=\left(\begin{array}{cccc}\psi_1&-\psi_2^*&\psi_3&\psi_4^*\\
                                         \psi_2&\psi_1^*&\psi_4&-\psi_3^*\\
                                         \psi_3&\psi_4^*&\psi_1&-\psi_2^*\\
                                         \psi_4&-\psi_3^*&\psi_2&\psi_1^*\end{array}\right)
\] 
is discussed in \cite{Hestenes86} and \cite{Launesto01}.

Penrose and Rindler\cite{PR86} discuss that the Dirac equation
\[
\gamma_{a\rho}^\sigma\nabla^a\psi^\rho=\hbar^{-1}m\psi^\sigma
\]
and the Dirac-Weyl equation 
\[
\gamma_{aR}^{S'}\nabla^a\psi^R=0,
\]
where $\gamma_{a\rho}^\sigma$ and $\gamma_{aR}^{S'}$ are operators to translate the 4 dimensional tangent vector field ${\bf V}^a$ to spinor form ${\bf S}^\sigma$ and ${\bf S}^R$, respectively, can be treated in the twistor space, where interaction of quaternions is treated. Although neutrino is not massless and does not satisfy the Dirac-Weyl equation, extension of the electromagnetic interaction to the electro-weak interecation including massive neutrinos is possible. 

When we identify the boson as a Higgs boson and compare the ATLAS experiment of decay into two jet photons in which one photon mass is about 100GeV\cite{ATLAS13}, which is close to the total energy of a $Z$ boson which makes a jet, and the energy of another photon becomes about 20GeV. We expect this photon becomes the source of 18 GeV photon from GRB 940217 which was difficult to produce from the inverse Compton scattering\cite{GRB08}.

\vskip 1 true cm
\centerline{\bf Acknowledgments}
\vskip 0.5 true cm
We thank Dr. Dos Santos for sending his thesis, which contains valuable information on the chaos via devil's staircase, and anonymous referees for giving us information on \cite{LVK10,LK13} and  \cite{GKD00,DGK03,DGKMS07}. 
 Thanks are also due to Dr. F. Nakamura of Wolfram Research for a support in programming using Mathematica 9.

\vskip 1 true cm


\begin{thebibliography}{99}
\bibitem{HGO93}  Hayes,S., Grebogi,C. and  Ott,E.[1993], Communicating with Chaos, 
Rev. Mod. Phys. {\bf 70}, 2031-2034.
\bibitem{EckRu85} Eckman,J.-P. and Ruelle,D.[1985],
Ergodic theory of chaos and strange attractors,  Rev. Mod. Phys. {\bf 57}, 617.
\bibitem{GrPr82} Grassberger,P. and Procaccia, I.[1983],
Measuring the Strangeness of Strange Attractors, Physica{\bf 9D}189-208.
\bibitem
{GOPY84} Grebogi,C., Ott,E., Pelikan,S. and Yorke,J.A.[1984], Strange Attractors that are not Chaotic, Physica {\bf 13D}, 261.
\bibitem{DGO89} Ding,M., Grebogi,C. and Ott, E.[1989],
Dimension of Strange Nonchaotic Attractors, Phys. Lett. A{\bf 137},167.
\bibitem{HeHa94} Heagy,J.F. and Hammel, S.M. [1994],
The birth of strange nonchaotic attractors, Physica {\bf D70} 140-153.
\bibitem{Kel96} Keller,G.[1996], A note on strange nonchaotic attractors,
Fundamenta Mathematicae {\bf 151}, 139.
\bibitem{Gle02} Grendinning,P.[2002], Global attractors of pinched skew products,
Dynamical Systems {\bf 17}, 287.
\bibitem{AlMi08} Alsed\`a,L. and Misiurewicz, M. [2008],
Attractors for unimodal quasiperiodically forced maps, 
Journal of Difference Equations and Applications, {\bf 14} 1175.
\bibitem{GrJa13}  Gr\"oger,M. and  J\"ager,T. [2013], Dimension of Attractors in Pinched Skew Products,
arXiv:1111.6574 v2[math.DS].
\bibitem{Chua71}  Chua,L.O.[1971], Memristor-The Missing Circuit Element,
IEEE Transactions on Circuit Theory, {\bf CT-18}, 507.
\bibitem{ChKa76}  Chua,L.O. and Kang,S.M.[1976], 
Memristive Devices and Systems, Proceedings of the IEEE, {\bf 64}, 209.
\bibitem{MuCh10} Muthuswamy,B. and Chua, L.O.[2010], Simplest Chaotic Circuit,
International Journal of Bifurcation and Chaos, {\bf 20}, 1567.
\bibitem{GiLe09}  Ginoux,J.-M. and Letellier, Ch.[2009],
Flow curvature manifolds for shaping chaotic attractors: I. R\"ossler-like systems,
J. Phys. A: Math. Theor. {\bf 42}, 285101-17.
\bibitem{GGJLF10} Gilmore, R., Ginoux,J-M., Jones,T., Letellier,C. and Freitas,U.S.[2010], 
Connecting curves for dynamical systems, J. Phys. A: Mathe Theor. {\bf 43},  255101-13.
\bibitem{GLC10} Ginoux,J.M., Letellier,Ch. and Chua,L.O.[2010], 
Topological Analysis of Chaotic Solution of Three-Element Memristive Circuit,
International Journal of Bifurcation and Chaos {\bf 20},3819-3827.
\bibitem{Chua11}  Chua,L.O.[2011], Resistance switching memories are memristors, Appl. Phys. {\bf A 102}, 765-783.
\bibitem{CKM86}  Chua,L.O., Komuro,M. and  Matsumoto,T.[1986], The Double Scroll Family,
IEEE Transactions on Circ.Syst, CAS{\bf 33},1073.
\bibitem{SaNa03} Saito,T. and Nakano,H.[2003], Chaotic Circuits Based on Dependent Switched Capacitors, CP676 Experimental Chaos, AIP Proceedings, 3-13.
\bibitem{PZC94} Pivka,L.,  Zheleznyak,A.L. and Chua,L.O.[1994], 
Arnold's Tongues, Devil's Staircase and Self-similarity in the Driven Chua's Circuit,
 International Journal of Bifurcation and Chaos {\bf 20},1743-1753.
\bibitem{CYY86}  Chua,L.O., Yao,Y. and Yang,Q.[1986], Devil's Staircase Route to Chaos in a Non-Linear Circuit, Circuit Theory and Applications, {\bf 14}, 315-329.
\bibitem{MTCP05} Manimehan,I., Thamilmaran,K.,  Chidambaram,G. and  Philominathan, P. [2005], Transition From Torus To Chaos In A Piecewise-linear Forced Parallel LCR Circuit, National Conference on Nonlinear Systems \& Dynamics, 1-4.
\bibitem{ZhLi97}  Zhu,Z. and  Liu,Z.[1997], 
Strange Nonchaotic Attractors of Chua's Circuit with Quasiperiodic Excitation,
International Journal of Bifurcation and Chaos, {\bf 7}, 227-238.
\bibitem{BaCa98} Baptista,M.S. and  Cardas,I.L.[1998] ,
Phase-Locking and Bifurcations of the Sinusoidally-Driven Double Scroll Circuit,
Nonlinear Dinamics {\bf 17}, 119-139.
\bibitem{ThLa02} Thamilmaran,K. and Lakshmanan,M. [2002], 
Classification of Bifurcations and Routes to Chaos in a Variant of Murali-Lakshmanan-Chua Circuit,  International Journal of Bifurcation and Chaos {\bf 12}, 783-813.
\bibitem{LVK10} Leonov,G.A., Vagaitsev,V.I. and Kuznetsov,N.V.[2010], Algorithm for Localizing Chua Attractors Based on the Harmonic Linearization Method, ISSN 1064-5624, Doklady Mathematics 2010, {\bf 82}, 663-666.
\bibitem{LK13} Leonov,G.A. and Kuznetsov,N.V.[2013], Hidden Attractors in Dynamical Systems, From Hidden Oscillators in Hilbert-Kolmogorov, Aizerman, and Kalman Problems to Hidden Chaotic Attractor in Chua Circuits, International Journal of Bifurcation and Chaos {\bf 23}, 1330002-1-69.
\bibitem{FuNi07} Furui,S. and Niiya,S.[2007], 
Shilnikov chaos control using homoclinic orbits and the Newhouse region,
Chaos, Solitons and Fractals, {\bf 34}, 966-988.
\bibitem{PF97} Pikovsky,A. and Feudel,U.[1997], Comment on "Strange nonchaotic attractors in autonomous and periodically driven system", Phys. Rev. E, {\bf 56}, 7320-7321.
\bibitem{Cartan66} Cartan,\'E.[1966],  {"The Theory of Spinors"}, Dover Pub. 
\bibitem{SF12} Furui,S.[2012], Fermion Flavors in Quaternion Basis and Infrared QCD, Few-Body Syst {\bf 52}, 171-187.
\bibitem{SF14} Furui,S.[2014a], Triality selection rules of Octonion and Quantum Mechanics, arXiv:1409.3761v2[hep-ph,hep-th].
\bibitem{Santos98} Dos Santos, S. and Planat M. [2000], Generation of $1/f$ Noise in Locked Systems Working in Nonlinear Mode, IEEE UFFC {\bf 47}, 1147.
\bibitem{SF13a} Furui,S.[2013a], Chaos in Electronic Circuits and Studies of Bifurcation Mechanisms, Proceedings of International Conference of Numerical Analysis and Applied Mathematics 2013(ICNAAM2013), Rhodos, AIP proceedings 1558, p.2233-2236.
\bibitem{SF13b} Furui,S.[2013b], Clifford Algebra and Physical and Engineering Sciences, Proceedings of International Conference of Numerical Analysis and Applied Mathematics 2013(ICNAAM2013), Rhodos, AIP proceedings 1558, p.521-524.
\bibitem{SF14b} Furui,S. [2014b] Axial anomaly and the triality symmetry of leptons and hadrons,  Few-Body Syst {\bf 55}, 1083-1097.
\bibitem{SF15} Furui,S. [2015] Cartan's Supersymmetry and the decay of a $H^0(0^+)$, arXiv:1504.03795[hep-ph].
\bibitem{FT14} Furui,S. and Takano,T. [2014] On the amplitude of external perturbation and chaos via devil's staircase in the Muthuswamy-Chua system,  To be published in IJBC,  arXiv:1406.4346[nlinCD].
\bibitem{Hestenes86} Hestenes,D. [1986] Clifford Algebra and the Interpretation of Quantum Mechanics, Chisholm, J.S.R. and Commons, A.K. (Eds) Clifford Algebra and their Appliations in Mathematical Physics, Reidel, Dordrecht/Boston, 321-346.
\bibitem{Launesto01} Launesto, P. [2001] {\it Clifford Algebras and Spinors} Second Edition,
Cambridge Univ. Press, Cambridge.
\bibitem{KKLN93} Khibnik, A.I., Kuznetsov, Y.A., Levitin, V.V., Nikolaev, E.V. [1993], Continuation techniques and interactive software for bifurcation analysis of ODEs and iterated maps,  Physika {\bf D62}, (1993), 360-371.
\bibitem{DBS78} De Boor, C. and Swartz, B. [1978], Collocation at Gaussian pints, SIAM K. Num. Anal., {\bf 10} 582-606.
\bibitem{GKD00} Govaerts, W., Kuznetsov,Y.A. and Dhooge,A. [2000], NumericalContinuation of Limit Cycle in MATLAB.
\bibitem{DGK03} Doedel, E.J., Govaerts,W. and Kuznetsov, Yu.A. [2003] SIAM J. Numer. Anal, {\bf 41} 401-435.
\bibitem{DGKMS07} Dhooge,A., Govaerts,W. , Kuznetsov,Yu.A., Meijer, H.G.E. and Sautois,B. [2007], New Features of the software MatCont for bifurcation analysis of dynamical system., Mathematical and Computer Modelling of Dynamical Systems.
\bibitem{PR86} Penrose, R. and Rindler, W. [1986] {\it Spinors and space-time} vol 2. Cambridge Univ. Press, Cambridge.
\bibitem{ATLAS13} The ATLAS collaboration [2013], Jet energy measurement with the ATLAS detector in proton-proton collisions at $\sqrt s=7$TeV , Eur. Phys. J. {\bf C 73}
 2304 ,@arXiv: 1112.6426 v1 [hep-ex].
\bibitem{GRB08} Fan, Yi-Z., Piran, T., Narayan,R. and Wei, Da-M. [2008], High-energy afterglow emission from gamma-ray bursts, Mon. Not. R.Astron.Soc.{\bf 384}.1483-1501.

\end{thebibliography}
\end{document}